\pdfoutput=1
\documentclass{ifacconf}

\usepackage[hyphens]{url}
\usepackage{amsmath, amsfonts, mathtools, amssymb, bm}
\usepackage{graphicx}      
\usepackage{natbib}        

\newcommand{\dif}{\,\operatorname{d}}
\newcommand*{\dpd}[3][]{\frac{\partial^{#1} #2}{\partial #3^{#1}}}
\newcommand*{\pd}[3][]{\frac{\partial^{#1} #2}{\partial #3^{#1}}}
\newcommand*{\del}[1]{\left(#1\right)}
\newcommand*{\sbr}[1]{\left[#1\right]}
\newcommand*{\cbr}[1]{\left\{#1\right\}}

\DeclareMathOperator{\expect}{\mathbb{E}}
\DeclareMathOperator{\probability}{\mathbb{P}}
\DeclareMathOperator{\cov}{\operatorname{cov}}
\DeclareMathOperator{\var}{\operatorname{var}}

\usepackage{etoolbox, comment, xcolor, mdframed}
\newtoggle{preprint}
\settoggle{preprint}{true}
\iftoggle{preprint}{%
   \usepackage[tracking=true, expansion, protrusion]{microtype}%
}{%
   \usepackage[tracking=true, expansion, protrusion]{microtype}%
}
\iftoggle{preprint}{%
   \newenvironment{skippedproof}[1]{%
      \begin{pf}{#1}%
   }{%
      \end{pf}%
   }%
}{\excludecomment{skippedproof}}

\begin{document}
\begin{frontmatter}
	\title{Estimation Sample Complexity of a Class of Nonlinear Continuous-time Systems\thanksref{footnoteinfo}}

	\thanks[footnoteinfo]{This work was supported by the National Science Foundation CAREER Program (Grant No.~2046292).}

	\author[First]{Simon Kuang}
	\author[First]{Xinfan Lin}

	\address[First]{University of California, Davis; Davis, CA 95616 (e-mail: slku@ucdavis.edu, lxflin@ucdavis.edu).}

	\begin{abstract}                
		We present a method of parameter estimation for large class of nonlinear systems, namely those in which the state consists of output derivatives and the flow is linear in the parameter.
		The method, which solves for the unknown parameter by directly inverting the dynamics using regularized linear regression,
		is based on new design and analysis ideas for differentiation filtering and regularized least squares.
		Combined in series, they yield a novel finite-sample bound on mean absolute error of estimation.
	\end{abstract}


\end{frontmatter}
\section{Introduction}
Most physical laws are continuous-time.
Examples include Newtonian motion \(\ddot x = m^{-1} F\), Fickian diffusion \(\dot q = D \Delta q\), and capacitance \(\dot v = C^{-1} i\).
These three relationships are semilinear (in fact, linear) ordinary differential equations in the state variables \(x\), \(q\), and \(v\); with vector fields that are linear in the parameters \(m^{-1}\), \(D\), and \(C^{-1}\), respectively.
We view the states as observable and the parameters as unknown.
System identification, as a way of measurement, takes on the emphasis of parameter estimation (as opposed to model prediction); this interpretation of system identification has universal application across engineering, e.g.~magnetic resonance imaging \citep{maidens_optimizing_2016} and battery diagnostics \citep{lin_modeling_2019}; and in other areas such as monetary policy econometrics \citep{chen_should_2023}.

We focus on the case in which all of the state variables are derivatives of the observed output.
In chaos-related literature, this structure is called a hyperjerk system.
In the geometric control literature, the observed output can also be understood as a flat output of a differentially flat system.
In the econometrics and control literature, this structure can be considered a nonlinear continuous-time autoregression with observation noise instead of the more customary process noise.
\section{Problem statement}
Let \(\Theta = \mathbb{R}^p\) be parameter space.
The output is \(y: [0, T] \times \Theta \to \mathbb{R}\),
and the state \(\xi \in \mathbb{R}^{m}\) consists of the derivatives of \(y\) of orders \(0\) through \(m - 1\).
The state dynamics are
\begin{align}
	\dpd[m]{}{t}
	y(t, \theta)
	 & =
	\theta^\intercal
	\phi(\xi(t, \theta))
	\label{eq:problem-model}
\end{align}
subject to the initial condition \(\xi(0, \theta) = \xi_0\), where \(\phi\) is some known function that does not depend on \(\theta\).

\begin{assum}
	\label{assumption:holder}
	The feature map \(\phi: \mathbb{R}^m \to \mathbb{R}^p\)  is \(\alpha\)-H\"older continuous with coefficient \(C_\phi\).\footnote{For a system with exogeneous input, the theory also works with a non-autonomous \(\phi\).}
\end{assum}
\begin{assum}
	\label{assumption:taylor-extra}
	\(\pd[m + 1]{}{t} y\) exists and is Riemann integrable on \([0, T]\).
\end{assum}


Let \((\Omega, \mathcal{F}, \mathbb{P})\) be a probability space with expectation \(\expect\), where \(\Omega\) is the sample space, \(\sigma\)-field \(\mathcal{F}\), and probability measure \(\mathbb{P}\).



We may observe \(y\) at \(n\) evenly spaced observations through the random variable \(z\) with the following law:
\begin{align}
	z_{i, n}
	 & = y(in^{-1} T, \theta) + w(in^{-1} T),
	\\
	i
	 & = 1, 2, \ldots, n.
\end{align}
\begin{assum}
	\label{assumption:noise-square-integrable}
	\(\expect w(t)^2\) is Riemann integrable in \(t\).
\end{assum}

\begin{defn}
	A \emph{discrete estimator} of \(\theta\) is a random variable \(\hat \theta : \Omega \to \Theta\) expressed as a function of \(\{z_{i, n}\}_{1 \leq i \leq n}\) that does not depend explicitly on \(\theta\).
\end{defn}




The goal is to find an estimator design that can be made provably accurate by increasing \(n\), which increases the quantity and frequency \(T/n\) of data collected.
\begin{prob}
\label{problem}
Find a sequence of estimators \(\{\hat \theta_n\}_{n \in \mathbb{N}}\) such that
 \(\hat \theta_n \overset{\mathbb{P}}{\longrightarrow } \theta  \), 
 with quantitative guarantees for finite \(n\).
\end{prob}




\iftoggle{preprint}{}{All but the most essential proof have been omitted.}

%
\section{Related work}




\subsection{Discretization does \emph{not} solve Problem~\ref{problem}}
The discrete-time model \(X_{i + 1} \approx f(X_i, \theta) \), where \(X\) is state, \(f\) is the dynamics, and \(\approx\) denotes process or observation noise, is a mainstay in system identification due to its resemblance to nonlinear regression \(Y \approx f(X, \theta)\).
There is recent progress in finite-sample identification of discrete-time linear \Citep{sarkar_finite_2021, tsiamis_statistical_2022-1, ziemann_tutorial_2023,matni_tutorial_2019} and parameter-linear \Citep{mania_active_2022} systems.
 Can discrete-time identification theory solve the continuous-time Problem~\ref{problem}? The answer is: not easily.
An overview of continuous-to-discrete challenges in system identification is available in \cite{ljung_issues_2010}.

One might argue that discretization converts continuous-time models to discrete-time models.
But \eqref{eq:problem-model} cannot be translated into discrete form without compromising either accuracy or parameter-linearity.
For instance, one can obtain the equivalent model
\begin{align}
	y((i+1)n^{-1}T, \theta) = f(y(i n^{-1} T, \theta), \theta) 
	\label{eq:equivalent-discrete-model}
\end{align}
where the dynamic map \(f\) is computed by solving an initial value problem (IVP).
But the right side of this equation is no longer linear in \(\theta\).

One could take a Taylor expansion of \eqref{eq:equivalent-discrete-model} for \(n^{-1} T \approx 0\) in order to preserve the parameter-linearity.
That would be equivalent to choosing a numerical scheme for integrating the IVP above, with the need to deal with the integration error.
This practice would be at least as hard as estimating differentiation error (our approach).




\subsection{Curve fitting does \emph{not} solve Problem~\ref{problem}}
In the prediction error method (PEM) for nonlinear system identification
(\Citealp{ljung_system_1999}; \Citealp{ljung_perspectives_2010}; \Citealp[sec.~5.2]{keesman_system_2011}),
we find \(\hat\theta\) such that the predicted flow generated by \(\hat \theta^\intercal \phi\) best fits the measured \(y\).
Some of the PEM literature is grounded in classical theory on general nonlinear least squares estimation \Citep{chen_strong_2020,wu_asymptotic_1981,lai_asymptotic_1994,jennrich_asymptotic_1969}
In certain large-sample limits, extremum estimators such as PEM and maximum likelihood estimation can be asymptotically consistent.
{
\setcitestyle{square}%
\begin{thm}[{%
	\Citet[Thm. 4.1.1]{amemiya_advanced_1985}}]
	Assume:
	\begin{enumerate}
		\item The parameter space \(\Theta\) is a compact subset of \(\mathbb{R}^p\).
		\item \(Q_n(\omega, \theta)\) is measurable in the outcome \(\omega \in \Omega\) and continuous in \(\theta \in \Theta\).\footnote{e.g.~a sum-of-squares error between random observation outcome \(\omega\) and prediction at \(\theta\).}
		\item \(n^{-1} Q_n(\omega, \theta) \to Q(\theta)\) in probability uniformly\footnote{
			i.e. \(\lim_{n \to \infty} \probability\sbr{\sup_{\theta\in\Theta} n^{-1}Q_n(\omega, \theta) < \epsilon} = 1\).
		} in \(\theta\) as \(n \to \infty\), and \(Q(\theta)\) has a unique \textbf{global} maximum at \(\theta_0\).
	\end{enumerate}	
	Let \(\hat \theta_n(\omega) \in \arg \max_{\theta} Q_n(\omega, \theta)\) be a measurable extremum.
	Then \(\hat \theta_n \overset{\mathbb{P}}{\to} \theta_0\).
\end{thm}
}
The global maximum estimator can be difficult to locate \Citep[p.~110]{amemiya_advanced_1985},
and we give such an example in Section~\ref{section:numerical-example} in comparison with our proposed estimator.
Furthermore, the global maximum estimator does not come with even asymptotic tail guarantees such as normality.
Asymptotic normality requires a number of higher regularity and rank conditions \Citep[Thm.~4.1.3]{amemiya_advanced_1985}, as well as 
the following result on \textbf{local} extrema: if \(Q_n\) has multiple local maxima, one of them converges in \(L^2(\Omega)\) to a local maxima of \(Q\) \Citep[Thm.~4.1.2]{amemiya_advanced_1985}.
On the other hand, our estimator comes with a moment guarantee valid at finite \(n\); and it is a ``plug-in'' estimator that does not require solving any implicit nonlinear equations.
Our estimator also works without continuity in \(\theta\); see Section~\ref{section:theoretical-examle}.





\subsection{Linear theory does \emph{not} solve Problem~\ref{problem}}
We have finite-sample guarantees when system identification amounts to estimating an operator \(\Theta\) in the linear relationship
\begin{align}
	Y & = \Theta X + \text{noise},
	\label{eq:parameter-linear-template}
\end{align}
where generally, \(X\) depends on the initial condition, system input, and observable states; \(Y\) is an output; and \(\Theta\) is an system representation such as a Hankel matrix or transfer function.
After multiplying both sides by a generalized inverse of \(X\), estimation error in \(\Theta\) is bounded in terms of \(X\) and the noise profile (\Citealp{ljung_system_1999, sarkar_finite_2021, pillonetto_regularized_2022,tsiamis_statistical_2022-1}; \Citealp[chap.~11]{pintelon_system_2012}; \Citealp[sec.~5.5.4.1]{bittanti_model_2019}).

We inherit an impressive legacy of theory and practice on identifying linear systems.
Linear system identification, for all practical purposes, has been a solved problem for decades.
But the conclusions mostly do not apply to the nonlinear system \eqref{eq:problem-model}, which does not satisfy scaling or superposition in \(y\).
That is why we have selected a nonlinear system, \eqref{eq:problem-model} which does not satisfy scaling or superposition in \(y\), and which enjoys no simplification under taking Laplace transforms.

\section{A differentiation-based approach}
We attack \eqref{eq:problem-model} directly by solving a linear regression for predictors \(\dpd[m]{}{t} y(t)\) in terms of regressors \(\phi(\xi(t, \theta))\).
This rewrites \eqref{eq:problem-model} into the form
\begin{gather}
	Y
	= \Theta \del{X + \text{error}_X} - \text{error}_Y,
	\label{eq:intro-eiv}
\end{gather}
differing from \eqref{eq:parameter-linear-template} only in the errors-in-variables \(\text{error}_X\).

To get the regressors \(\phi(\xi(t, \theta))\)
we must differentiate the measurements \Citep{soderstrom_least_1997, van_breugel_numerical_2020}.
Conventional wisdom holds that differentiation of noisy data is plagued by bias (due to aliasing high-frequency signal) and fluctuation (due to amplification of high-frequency noise).
But this impression is fallible on both counts---bias and fluctuation.

\textbf{Concerning bias}: any derivative filter introduces bias in the form of generalized aliasing.
Even without noise, it is never possible to get perfect derivatives for a general smooth \(y\), because one cannot with certainty impute what happens in between samples.
But suppose, as is often the the case, that the sampling period \(\Delta t\) is small compared to the regularity of \(y\).
When applying PEM to such a case, it would seem viable to integrate the flow using a linear multi-step scheme\footnote{A good exposition can be found in \cite{dattani_linear_2008}. \cite{butcher_general_2006} is a comprehensive treatment.} with steps aligned to the measurement schedule.
The quadrature error is controlled by comparing \(y\) to local polynomial approximants.
Our differentiator will share the local polynomial approximation, but we manage the error systematically in Lemma~\ref{lemma:diff-bias} and beyond.

\textbf{Concerning fluctuation}: any derivative filter amplifies noise, as the ideal LTI differentiator has Laplace transfer function \(s\).
Our regressor matrix is noisy, but we manage this variability with ridge regularization in Appendix~\ref{section:noisy-diff}.\footnote{Appendices are only available in the preprint version of this article at \Citet*{kuang_estimation_2024}.}
To compare with PEM:
consider estimation of \(\dot y = \theta\).
The time average of \(\hat {\dot y}\), which is the differentiation-based estimate, is very close to the least-squares PEM estimate.


\subsection{Contributions}


\textbf{An arbitrary-order finite difference filter for differentiating noisy signals}:
	      with modest regularity assumptions on trajectory and noise, we are able to bound the mean squared error of the derivative estimates.
	      The filter resembles Savitzky-Golay \Citep{schmid_why_2022,niedzwiecki_application_2021,ochieng_adaptive_2023} and orthogonal polynomial filtering \Citep{othmane_contributions_2022,diekema_differentiation_2012}, but emphasizes the bias-variance tradeoff when filtering noisy data, following \cite{sadeghi_window_2020,krishnan_selection_2013}.

After differentiation, our method uses the derivative estimates to invert \eqref{eq:problem-model}.
Our use of finite-difference filters to estimate a continuous-time system resembles \Citet{soderstrom_least_1997}, which treats of a linear system with process noise, but our analysis makes the quantitative bias-variance tradeoff explicit.
The concept of inverting linear equations to estimate a continuous-time dynamic parameter has a spectral interpretation by \Citet{unbehauen_identification_1997}.

\textbf{A novel analysis of ridge regression subject to square-integrable error in both the regressors and predictors.}
Noise in both variables is called \emph{errors-in-variables} in statistics (\Citealp{gleser_limiting_1987,hirshberg_least_2021,zhang_asymptotic_2020}; \Citet[sec~8.2]{pintelon_system_2012})
and system identification
\Citep{soderstrom_errors--variables_2018, barbieri_recursive_2021,khorasani_non-asymptotic_2021,fosson_concave_2021}.
These problems are ill-behaved because they require the inversion of a random matrix which could \emph{a priori} be singular with positive probability, leading to heavy tails.
Additionally, the phenomenon of regression dilution states that the inverse of a random matrix is too small on average.
We account for both of these issues in our analysis.

In brief, let \(\sigma > 0\) be an unreliable singular value in the data.
Ridge regularization picks \(\lambda > 0\) to control \(\del{\sigma + \lambda}^{-1}\).
The na\"ive bound is \(\del{\sigma + \lambda}^{-1} \leq \lambda^{-1}\) which is an \(L^\infty\)-type bound that is forced by the worst-case scenario \(\sigma = 0\).
Rather, we use a very detailed \(L^1\)-type pseudoinverse perturbation bound, Lemma~\ref{lemma:head-body-tail} (Appendix~\ref{section:noisy-diff}).
This is the main idea in the proof of Thm~\ref{thm:mae-after-lambda}.

\section{Preliminaries}
Unless otherwise specified, \(\left\|\cdot \right\|\) denotes the 2-norm of a vector and the operator 2-norm of a matrix.
We write \(\min(x, y)\) as \(x \wedge y\).
We write \(\max(0, x)\) as \(x^+\).
The least singular value of a matrix \(A\) is \(\sigma_\text{min} (A)\).

We say that a function is Riemann integrable if its equally-spaced Riemann sums converge.

\section{Solution to Problem~\ref{problem}}
This section constructs an estimator that solves Problem~\ref{problem}.
\subsection{Differentiation}
Partition the \(n\) random variables \(z_{j}\)
	into to \(n'\) windows, each containing
\(N\) consecutive samples, where \(N\) will be determined later.
The windows can be chosen to be disjoint, \(n' = n/N\); overlapping by \(n-1\) points, \(n' = n-N+1\); or anything in between.
\footnote{Sharpening the analysis to cover overlapping-window designs is an area of ongoing work.}

Let \(n_i\) be the first index of the \(i\)th window.
Then the start time of the \(i\)th window is \(t_i = n_in^{-1} T\).
Augment \(\xi\) of \eqref{eq:problem-model} with an extra derivative to define \(x^d_i = \pd[d]{}{t} y(t_i)\), \(d = 0, 1, \ldots, m\).
Regard \(x_i\) as an \(\mathbb{R}^{m + 1}\) vector.

\begin{thm}
	\label{thm:differentiation}
	For some choice of \(N\), there exist statistics \(\hat x_i^d\) for all \(i \in [1\ldots n']\) and \(d \in [0\ldots m]\),
    given as linear functions of the data \(z_j\), \(j \in [1\ldots n]\),
	that satisfy
	\begin{align*}
		\expect
		\sum_{i = 1}^{n'}
		\sum_{d = 0}^{m}
		\left\| \hat x_i^d  - x_i^d\right\|^2
		& \leq
		\epsilon
		n'
		n^{-\frac{2}{2m + 3}},
	\end{align*}
	where \(\epsilon > 0\) is independent of \(n\).
\end{thm}
This counterintuitive result on tuning arbitary-order finite-difference filters for noisy data may be the first of its kind suggests that noisy data can be practically differentiated for statistical use in a certain large-sample limit.

The proof using orthogonal polynomials is constructive and given in Appendix~\ref{section:noisy-diff}.
Henceforth the notation \(\hat x_i^d\) will refer to the estimators provided by Theorem~\ref{thm:differentiation}.

\subsection{Regularized least squares}
Let \(P\) be the projection matrix onto the first \(m\) entries.
Introduce the shorthand \(\phi_i = \phi(Px_i)\) and
\(\hat \phi_i = \phi(P\hat x_i)\).

Let \(\Phi \in \mathbb{R}^{n' \times p}\)
be the matrix having \(\phi_i^\intercal, i = 1, 2, \ldots, n'\) as rows, and let \(\hat \Phi\) be the same, with \(\hat \phi_i\) replacing \(\phi_i\).
Let \(\lambda \in (0, \sigma_\text{min}(\Phi))\) be a regularization coefficient to be determined later.
Define \(R, \hat  R\in \mathbb{R}^{(p + n') \times p}\) as the block matrices
\begin{align}
	R & = \begin{pmatrix}
		      0_{p\times p} \\
		      \Phi
	      \end{pmatrix},
	\quad \hat R = \begin{pmatrix}
		               \lambda I_{p} \\
		               \hat \Phi
	               \end{pmatrix}.
\end{align}
Define \(u, \hat u  \in \mathbb{R}^{p + n'}\)  as
\begin{align}
	u & = \begin{pmatrix}
		      0_{p \times 1} \\
		      x^{m}_1        \\
		      x^{m}_2        \\
		      \vdots         \\
		      x^{m}_{n'}
	      \end{pmatrix},
	\quad \hat u = \begin{pmatrix}
		               0_{p \times 1} \\
		               \hat x^{m}_1   \\
		               \hat x^{m}_2   \\
		                \vdots    \\
		               \hat x^{m}_{n'}
	               \end{pmatrix}.
\end{align}

The system dynamics can be stated as \(u = R \theta\).
Assuming that \(R\) has rank \(m\), let \(R^\dagger\) be the Moore-Penrose pseudoinverse of \(R\), leading to the identity
\begin{align*}
	\theta      & = R^\dagger u
	\intertext{which motivates the estimator}
	\hat \theta & = \hat R^\dagger \hat u.
\end{align*}

\begin{thm}
    \label{thm:mae-after-lambda}
	For some choice \(\lambda = \lambda(n)\),
	\(\expect \left\|\hat \theta - \theta\right\| = O(n^{-\alpha/(2m +3)})\)
	as \(n\to \infty\).
\end{thm}
This result is the first finite-sample frequentist guarantee for parameter estimation of a large class of nonlinear continuous-time systems from discrete data.

The proof is constructive and given in Appendix~\ref{section:estimator}.

\section{Numerical example}
\label{section:numerical-example}
Consider the problem of estimating the coefficients \(\theta = (\theta_1, \theta_2)\) in the Van der Pol oscillator:
\begin{align*}
	\ddot x(t) &= \theta_1 (1 - x(t)^2) \dot x(t) + \theta_2 x(t),
    \quad 0 \leq t \leq T
    \intertext{The measurements are i.i.d.~Normal random variables}
    z_i &\sim \mathcal{N}(x(in^{-1} T), \sigma^2).
\end{align*}
Constants pertaining to the problem and the estimator are available in Table~\ref{fig:numerical-values}.

\begin{figure}
    \centering
    \includegraphics[width=\linewidth]{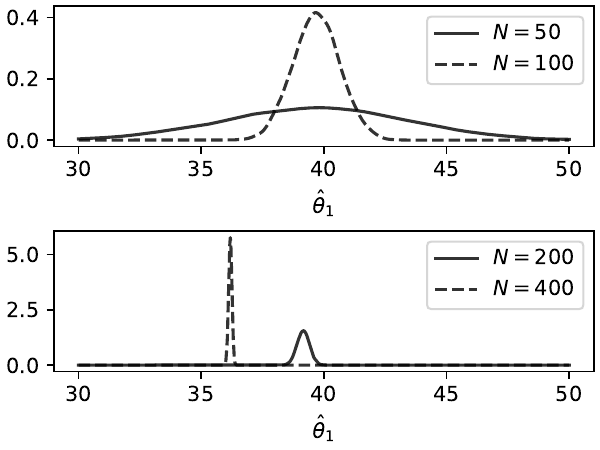}
    \caption{\label{fig:sampling-distribution-1} Sampling distribution of \(\hat \theta_1\) at different choices of \(N\).}
\end{figure}

\begin{figure}
    \centering
    \includegraphics[width=\linewidth]{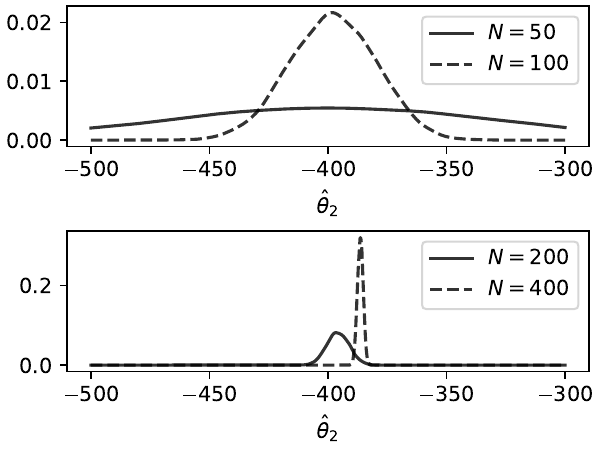}
    \caption{\label{fig:sampling-distribution-2} Sampling distribution of \(\hat \theta_2\) at different choices of \(N\).}
\end{figure}

For each of four different values of \(N\), 
we constructed an estimator \(\hat \theta\) using the filtered least squares idea (with disjoint windows) presented in this paper and used 10000 trials to estimate its sampling distribution.
The marginal distributions for \(\theta_1\) and \(\theta_2\) are presented in Figures~\ref{fig:sampling-distribution-1} and \ref{fig:sampling-distribution-2}, respectively.

We see that a higher value of \(N\), which corresponds to more aggressive smoothing, decreases the estimator's variance while increasing its bias.
The \(L^2\) error is also minimized at \(N = 200\), as we can see from absolute (Table~\ref{table:mse}) and relative errors (Table~\ref{table:rel-mse}).

\begin{figure}
    \centering
    \includegraphics[width=\linewidth]{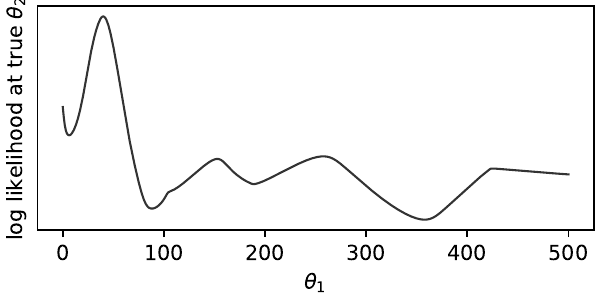}
    \caption{\label{fig:log-likelihood-1} Log-likelihood of \(\theta_1\) at the true \(\theta_2\).}
\end{figure}

\begin{figure}
    \centering
    \includegraphics[width=\linewidth]{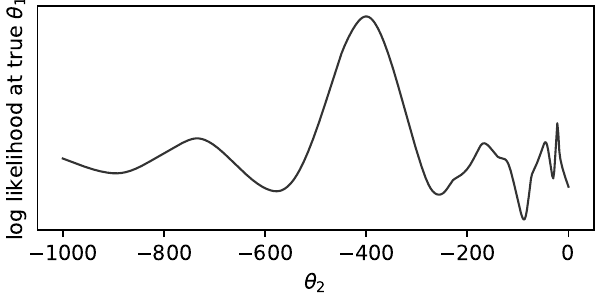}
    \caption{\label{fig:log-likelihood-2} Log-likelihood of \(\theta_2\) at the true \(\theta_1\).}
\end{figure}

In order to illustrate why a direct method such as our \(\hat \theta\) is apt for this problem, we point to the fact that the negative least squares objective, which corresponds to maximum likelihood, is two-dimensional and far from convex, which renders PEM practically unusable in this case.
Using the noisy data from one trial, we computed the likelihood function \(\mathcal{L}(\theta)\) and plotted it with \(\theta_2\) and \(\theta_1\) held constant at their true values in Figures~\ref{fig:log-likelihood-1} and \ref{fig:log-likelihood-2}, respectively.
These figures show that while the likelihood function attains a maximum at the true parameter value, the likelihood landscape is rife with local maxima, and maximum likelihood estimation is unreliable without strong prior information.




It is a promising two-step estimation scheme to compute \(\hat \theta\) via our plug-in estimator, then use the rough estimate to initialize a local search algorithm for the maximum likelihood or maximum a posteriori point estimate.

\begin{table}
    \centering
    \begin{tabular}{rrrr}
        \(N\)
        & \(\text{RMSE}(\hat \theta_1)\)
        & \(\text{RMSE}(\hat \theta_2)\)
        & \(\text{RMSE}((\hat \theta_1, \hat \theta_2))\)
        \\\hline
        50
        & 3.77
        & 72.74
        & 5305.53 \\
        100
        & 1.00
        & 18.45
        & 341.26 \\
        200
        & 0.88
        & 6.37
        & 41.37 \\
        400
        & 3.82
        & 13.69
        & 202.03
    \end{tabular}
    \caption{
        \label{table:mse}
        Experimentally determined root mean squared error (RMSE) of \(\hat\theta\) at different values of \(N\).
    }
\end{table}

\begin{table}
    \centering
    \begin{tabular}{rrrr}
        \(N\)
        & \(\text{\%RRMSE}(\hat \theta_1)\)
        & \(\text{\%RRMSE}(\hat \theta_2)\)
        & \(\text{\%RRMSE}((\hat \theta_1, \hat \theta_2))\)
        \\\hline
        50
        & 9.43
        & 18.19 
        & 18.12 \\
        100
        & 2.50
        & 4.61 
        & 4.60 \\
        200
        & 2.20
        & 1.59 
        & 1.60 \\
        400
        & 9.54
        & 3.42 
        & 3.54
    \end{tabular}
    \caption{
        \label{table:rel-mse}
        Relative root mean squared error (RMSE) of \(\hat\theta\) at different values of \(N\), obtained by normalizing columns of Table~\ref{table:mse} by \(|\theta_1|\), \(|\theta_2|\), and \(\left\|\theta\right\|\), respectively.
    }
\end{table}

\begin{table}
    \centering
    \begin{tabular}{ccr}
        Constant & Meaning & Value
        \\\hline
        \(\theta_1\) & parameter (nonlinear damping) & 40
        \\
        \(\theta_2\) & parameter (squared frequency) & -400
        \\
        \(T\) & end time & 1
        \\
        \(x(0)\) & initial displacement & 1
        \\
        \(\dot x(0)\) & initial rate & 20
        \\
        \(\sigma^2\) & observation variance & \(10^{-4}\)
        \\ 
        \(n\) & number of samples & 10000
        \\
        \(N\) & window size & 50, 100, 200, 400
        \\
        \(\lambda\) & regularization coefficient & 1
        \\
    \end{tabular}
    \caption{\label{fig:numerical-values} Values of constants in simulating a Van der Pol oscillator and estimating its parameter.}
\end{table}
\section{Theoretical example}
\label{section:theoretical-examle}
Let the system model be any instantiation of \eqref{eq:problem-model} with \(\phi\) satisfying \(\phi(0) = 0\), \(\nabla \phi(0) \neq 0\):
\begin{align}
	\dpd[m]{}{t}
	y(t, \theta)
	 & =
	\theta^\intercal
	\phi(\xi(t, \theta))
    \intertext{Let \(t_i\) be the observation times. Let the observations obey}
    z_i
    &= y(t_i) + w_i,
    \\
    \intertext{where the \(w_i\) are independently distributed as}
    w_i
    &= 
    \begin{cases}
        e + y(t_i) &\text{w.p. } 1/2
        \\
        -e -y(t_i) &\text{w.p. } 1/2
    \end{cases}
\end{align}
for some \(e \in \mathbb{R}\).
Because the noise \(w_i\) is an atomic random variable that depends on \(\theta\), a likelihood function does not exist; any solution has to rely on smoothing.
Our method, with its accuracy guarantees, still applies, because \(\expect \left\|w_i\right\|^2 = \left\|e + y(t_i)\right\|^2 < \infty\).

This example also shows the need for ridge regularization.
For all sufficiently small \(\epsilon\), with probability \(2^{-n}\), every row in the regressor matrix \(\hat\Phi\) will be equal to the same small vector of magnitude \(|e|\).
Accordingly, the unregularized pseudoinverse of \(\hat \Phi\) will have a singular value on the order of \(|e|^{-1}\), which can cause \(\hat \theta\) to blow up.
Such an event, in which \(\theta\) is unidentifiable, is exponentially unlikely.
Ridge regularization serves as a kind of insurance that truncates this tail risk, so that the estimation error is at least finite.
\section{Conclusion}
Continuous-time parameter-linear systems are widely useful, but they do not have an estimation theory with provable finite-sample accuracy.
While smoothed least squares is a mainstay in \emph{linear} system identification, we have contributed a theoretical basis for optimal smoothing and regularization in a \emph{nonlinear} problem in which traditional proof techniques centered on transfer function algebra do not apply.
By using the parameter-linear structure to reformulate the parameter estimation problem as linear least squares, we are able to use a stochastic continuity argument to prove \(L^1\) consistency in the limit as \(n \to\infty\) with \(T\) held constant.
This approach even works for problems with discontinuous observation noise in which maximum-likelihood large-sample theory does not apply.

Theorem~\ref{thm:mae-after-lambda} is likely the first identification result in the literature with
\begin{enumerate}
    \item quantitative, finite-sample consistency, for
    \item a large class of nonlinear systems \eqref{eq:problem-model} with continuous and unbounded parameter,
    \item observed at discrete times, with
    \item heavy-tailed noise.
\end{enumerate}
The mathematical technology underlying this result is interesting in its own right, and contains new methods for analyzing differentiation filters and regularized least squares.


\bibliography{export.bib}             
\section*{Appendices}
Appendices A through C are available in \Citet{kuang_estimation_2024}, the preprint version of this paper, available: \url{https://arxiv.org/abs/2312.05382}.








\iftoggle{preprint}{\clearpage}{}
\appendix
\iftoggle{preprint}{%
	\section{Proof of Theorem \ref{thm:differentiation}}
\label{section:noisy-diff}

\subsection{Building blocks of the filter}
Fix \(\rho: [0, 1] \to (0, \infty)\), a Riemann integrable function.
Define an inner product on the polynomials of degree at most\footnote{This is a rank-\(N\) quadratic form and therefore doesn't extend to an inner product on general \(f\) and \(g\).} \(N - 1\):
\begin{align*}
	\left\langle
	f, g
	\right\rangle_N
	 & = \frac{1}{N}
	\sum_{j = 1}^N
	\rho\del{\frac{j}{N}}
	f\del{\frac{j}{N}}
	g\del{\frac{j}N}.
\end{align*}
Let the polynomials
\(\{p^d_{N} : 0 \leq d \leq N-1\}\),
\(\deg p^d_N = d\) be an orthogonal family for each \(N\):
\begin{align*}
	\left\langle
	p^d_N, p^{d'}_N
	\right\rangle_N
	 & = 0
	\quad\text{when}\quad d\neq d'.
\end{align*}
As \(\left\langle
f, g
\right\rangle_N\)
converges to a Riemann integral as \(N \to \infty\), so \(p_N^d\) converges too.
Define the filter-specific constants
\begin{align*}
	h_N^d
	 & =
	\frac{1}{N}
	\sum_{j = 1}^N
	\rho\del{\frac{j}{N}}
	p_N^d\del{\frac{j}{N}}
	\del{\frac{j}{N}}^d \\
	\intertext{and}
	g_N^d
	 & =
	\frac{1}{N}
	\sum_{j = 1}^N
	\rho\del{\frac{j}{N}}
	\left|
	p_N^d\del{\frac{j}{N}}
	\right|.
	\notag
\end{align*}

Define the problem-specific constants
\begin{align*}
	M_{i}^d
	 & = \sup_{t_i \leq t \leq t_i + n^{-1}NT}
	\left|
	\dpd[d + 1]{}{t}y^{(d + 1)}(t)
	\right|
	\intertext{and}
	(s_{i}^d)^2
	 & =
	\frac{1}{N}
	\sum_{j = 1}^N
	\rho\del{\frac{j}{N}}^2
	p_N^d\del{\frac{j}{N}}^2
	\var w((n_i+j) n^{-1} T).
\end{align*}

\subsection{Choice of filter coefficients}
The filter is given by
\begin{align}
	\hat x^d_i
	 & = C
	\sum_{j =1}^{N}
	\rho\del{\frac{j}{N}}
	p^d_N\del{\frac{j}{N}}
	z_{n_i + j, n},
	\label{eq:diff-definition}
	\intertext{where}
	C
	 & = \frac{
		d! n^d
	}{
		N^{d + 1}
		T^d h_N^d
	}.
	\notag
\end{align}

Let \(\Delta x_i = \hat x_i - x(t_i)\).
\begin{lem}[Bias of the estimated derivative]
	\label{lemma:diff-bias}
	Let \(\hat x^d_i\) be as in \eqref{eq:diff-definition}.
	Then
	\begin{align*}
		\left|
		\expect \Delta x^d_i
		\right|
		 & \leq
		\del{\frac{M_i^dT}{d + 1}}
		\del{\frac{N}{n}}
		\frac{
			g_N^d
		}{
			h_N^d
		}.
	\end{align*}
\end{lem}
\begin{skippedproof}
	By Taylor's theorem,
	\begin{align}
		\expect z_{n_i + j, n}
		 & =
		y\del{\frac{(n_i + j)T}{n}}
		\notag
		\\
		 & = \sum_{k = 0}^d \frac{
			y^{(k)}(t_i)
		}{
			k!
		}
		\del{
			\frac{NT}{n}
		}^k
		\del{\frac{j}{N}}^k + R
		\label{eq:expect_z_i}
		\intertext{where the remainder \(R = R(i, d, n, N)\) satisfies}
		\left|R\right|
		 & \leq
		R_\text{max}
		=
		\del{
			\frac{NT}{n}
		}^{d+1}
		\frac{M_i^d}{(d + 1)!}.
	\end{align}
	We now insert this into \eqref{eq:diff-definition}.
	By the orthogonality property of \(p^d_N\), all of the lower-order Taylor expansion terms of \(\expect z_{n_i + j, n}\) vanish.
	\begin{align}
		\expect \hat x^d_i
		 & =
		C \sum_{j = 1}^N
		\rho\del{\frac{j}{N}}
		p_N^d\del{\frac{j}{N}}
		\frac{x^d_i}{d!}
		\del{\frac{NT}{n}}^d
		\del{\frac{j}{N}}^d \notag \\
		 & \phantom{=} +
		C \sum_{j = 1}^N
		\rho\del{\frac{j}{N}}
		p_N^d\del{\frac{j}{N}}
		R(j, d, n, N)
	\end{align}
	By our choice of \(C\), the first sum evaluates to \(x^d_i\).
	\begin{align}
		\left|
		\expect \hat x^d_i - x^d_i
		\right|
		 & \leq
		C N R_\text{max} g_N^d
		\notag
		\\
		 & =
		\del{\frac{M_i^dT}{d + 1}}
		\del{\frac{N}{n}}
		\frac{
			g_N^d
		}{
			h_N^d
		}.\notag
	\end{align}
\end{skippedproof}

\begin{lem}[Variance of the estimated derivative]
	\label{lemma:diff-variance}
	Let \(\hat x^d_i\) be as in \eqref{eq:diff-definition}.
	Then
	\begin{align*}
		\var \hat x^d_i
		 & \leq
		\frac{n^{2d}}{
			N^{2d + 1}
		}
		\del{
			\frac{
				d!  s_{i, N}^d
			}{
				T^{d} h_N^d
			}
		}^2.
	\end{align*}
\end{lem}
\begin{skippedproof}
	Direct calculation beginning with
	\begin{gather*}
		\begin{split}
			\var
			\hat x^d_i
			=
			C^2
			\sum_{j = 1}^N
			\Bigg[
				&\ \rho\del{\frac{j}{N}}^2
				p_N^d\del{\frac{j}{N}}^2\\
				&\quad\cdot \var w((n_i + j)n^{-1} T)\Bigg].
		\end{split}
	\end{gather*}
\end{skippedproof}

\begin{thm}[MSE of the estimated derivative]
	\label{thm:mse-d-i}
	The mean squared error (MSE) of \(\hat x_i^d\) satisfies
	\begin{align*}
		\expect \left\|\Delta x^d_i\right\|^2
		          & \leq  A_{i, N}^d \frac{N^2}{n^2}
		+  B_{i, N}^d \frac{n^{2d}}{N^{2d + 1}},
		\intertext{with constants}
		A_{i,N}^d & = \del{\frac{M_i^dT}{d + 1}}^2
		\del{
			\frac{
				g_N^d
			}{
				h_N^d
			}
		}^2
		\quad \text{and}
		\\
		B_{i,N}^d & = \del{
			\frac{
				d! s_{i,N}^d
			}{
				T^{d} h_N^d
			}
		}^2.
	\end{align*}
\end{thm}
\begin{pf}
	Follows from applying the identity \(\text{MSE} = \text{Bias}^2 + \text{Variance}\) to Lemmas~\ref{lemma:diff-bias} and \ref{lemma:diff-variance}.
\end{pf}

\subsection{Choice of \(N\)}
To pick the filter window size \(N\), add up Thm.~\ref{thm:mse-d-i} across all \(d = 0, 1, \ldots, m\).
\begin{align}
	\expect
	\left\| \Delta x_i \right\|^2
	  & \leq
	\sum_{d = 0}^m
	A^d_{i, N} \frac{N^2}{n^2}
	+
	\sum_{d = 0}^m
	B^d_{i, N} \frac{n^{2d}}{N^{2d + 1}}
	\\
	\intertext{We can then bound the total MSE across all windows as}
	\expect
	\sum_{i = 1}^{n'}
	\left\| \Delta x_i \right\|^2
	  & \leq
	\tilde A_N
	\frac{N^2}{n^2}
	+ \tilde B_N \frac{n^{2m}}{N^{2m + 1}},
	\\
	\tilde A_N
	  & =
	\sum_{i = 1}^{n'}
	\sum_{d = 0}^m
	A^d_{i, N},
	\\
	\tilde B_N
	  & =
	\sum_{i = 1}^{n'}
	\sum_{d = 0}^m
	B^d_{i, N}
	.
	\intertext{Optimizing in \(N\), we pick}
	N & =
	\sbr{
		\frac{(2m + 1)\hat B_N }{2 \hat A_N}
	}^{\frac{1}{2m + 3}}
	n^{\frac{2m + 2}{2m + 3}},
	\notag
\end{align}
so that the total MSE obeys
\begin{align}
	\expect
	\sum_{i = 1}^{n'}
	\left\| \Delta x_i  \right\|^2
	 & \leq
	\epsilon_N
	n'
	n^{-\frac{2}{2m + 3}},
	\label{eq:total-MSE}
\end{align}
where
\begin{align}
	\epsilon_N
	 & =
	2
	(2m + 1)^{\frac{1}{2m + 3}}
	\hat A_N^{\frac{2m + 1}{2m + 3}}
	\hat B_N^{\frac{2}{2m + 3}} (n')^{-1}
\end{align}
tends to a limit as \(N \to \infty\).

Likewise, let \(\epsilon_{N, m} \leq \epsilon_{N}\) be a constant such that
\begin{align}
	\expect
	\sum_{i = 1}^{n'}
	\left\| \Delta x^m_i  \right\|^2
	 & \leq
	\epsilon_{N, m}
	n'
	n^{-\frac{2}{2m + 3}}.
	\label{eq:highest-MSE}
\end{align}

	\section{Proof of Theorem \ref{thm:mae-after-lambda}}
\label{section:estimator}

Let
\(\Delta \phi_i = \hat \phi_i - \phi_i\).
By H\"older continuity,
\(\left\|\Delta \phi_i\right\| \leq C_\phi \left\|\Delta x_i\right\|^\alpha\).
Let \(\delta = \left\|\Delta \Phi\right\|\).
\subsection{Error analysis}
To analyze the error \(\Delta \theta = \hat \theta - \theta\),
let \(v = \hat u - u\) and \(S = \hat R^\dagger - R^\dagger\).
\begin{align}
	\Delta \theta
	 & = \del{R^\dagger + S} \del{u + v} - R^\dagger u
	\\
	 & = R^\dagger v + S\del{u + v}
	\\
	\left\|\Delta \theta\right\|
	 & \leq \frac{\left\|v\right\|}{\sigma_\text{min} (R)}
	+ \left\|S\right\| \left\|u\right\| +
	\left\|S\right\| \left\|v\right\|
	\label{eq:Delta-theta}
\end{align}
To extract the dependence on \(n'\), we introduce the variables
\begin{align}
	\bar\sigma
	 & = (n')^{-1/2}\sigma_\text{min}(\Phi) , \\
	\bar U
	 & = (n')^{-1/2} \left\|u\right\|
\end{align}
These are scaled so that they approach limits as \(n \to \infty\).

\begin{lem}[Moment of \(v\)]
	\label{lemma:v-moment}
	For all \(r \in (0, 2]\), the random variable \(v\) satisfies
	\begin{align*}
		\expect \left\|v\right\|^r
		 & \leq
		\frac{
			\epsilon_{N, m}^{r/2}
			(n')^{r/2}
		}
		{
			n^{\frac{r}{2m + 3}}
		}.
	\end{align*}
\end{lem}
\begin{skippedproof}
	By \eqref{eq:highest-MSE},
	\begin{align}
		\expect \left\|v\right\|^2
		 & = \expect \sum_{i = 1}^{n'}
		\left\|\Delta x_i^{m }\right\|^2
		\notag
		\\
		 & \leq
		\epsilon_{N, m}
		n'
		n^{-\frac{2}{2m + 3}}.
		\\
		\intertext{For all \(0 < r \leq 2\), Jensen's gives}
		\expect
		\left\|v\right\|^{r}
		 & \leq
		\frac{
			\epsilon_{N, m}^{r/2}
			(n')^{r/2}
		}
		{
			n^{\frac{r}{2m + 3}}
		}
		\label{eq:v-moment}.
	\end{align}
\end{skippedproof}

\begin{lem}[Moment of \(\delta\)]
	\label{lemma:delta-moment}
	The random variable \(\delta\) satisfies
	\begin{align*}
		\expect \delta^{2/\alpha}
		 & \leq
		\frac{
			C_\phi^{2/\alpha}
			\epsilon_N
			(n')^{1/\alpha}
		}{
			n^{\frac{2}{2m + 3}}
		}.
	\end{align*}
\end{lem}
\begin{skippedproof}
	Bounding \(\delta\) by the Frobenius norm of \(\Delta \Phi\), we have
	\begin{align}
		\delta^2
		 & \leq \sum_{i = 1}^{n'}
		\left\|\Delta \phi_i\right\|^2        \\
		 & \leq
		C_\phi^2
		\sum_{i = 1}^{n'}
		\left\|\Delta x_i\right\|^{2\alpha}   \\
		\intertext{By H\"older's inequality with exponents \(1/\alpha\) and \(1/(1 - \alpha)\),}
		 & \leq
		C_\phi^2
		\del{
		\sum_{i = 1}^{n'}
		\del{\left\|\Delta x_i\right\|^{2\alpha}}^{1/\alpha}
		}^{\alpha}
		\del{
			\sum_{i = 1}^{n'}
			1^{\frac{1}{1 - \alpha}}
		}^{1 - \alpha}
		\\
		\intertext{Raising both sides to the power \(1/\alpha\) and taking expectations,}
		\delta^{2/\alpha}
		 & \leq
		C_\phi^{2/\alpha}
		(n')^{1/\alpha - 1}
		\sum_{i = 1}^{n'}
		\left\|\Delta x_i\right\|^{2}         \\
		\expect \delta^{2/\alpha}
		 & \leq
		C_\phi^{2/\alpha}
		(n')^{1/\alpha - 1}
		\sum_{i = 1}^{n'}
		\expect \left\|\Delta x_i\right\|^{2} \\
		\intertext{Applying \eqref{eq:total-MSE},}
		 & \leq
		\frac{
			C_\phi^{2/\alpha}
			\epsilon_N
			(n')^{1/\alpha}
		}{
			n^{\frac{2}{2m + 3}}
		}.
	\end{align}
\end{skippedproof}

To control \(S\), we state and then apply a self-contained technical lemma%
\iftoggle{preprint}{, proven in Appendix~\ref{appendix-proof:perturbed-tikhonov-pseudoinverse}}{}.
\begin{lem}
	\label{lemma:perturbed-tikhonov-pseudoinverse}
	Let \(A \in \mathbb{R}^{m\times n}\) have full column rank and minimal singular value \(\sigma\), and let \(D\) be a real matrix of the same dimensions as \(A\),
	having operator norm \(\delta\).
	For any \(\lambda> 0\), let
	\begin{align*}
		B
		 & = \begin{pmatrix}
			     0_{n\times n} \\ A
		     \end{pmatrix}^\dagger,
		\quad
		\hat B
		=
		\begin{pmatrix}
			\lambda I_n \\ A + D
		\end{pmatrix}^\dagger
		, \quad \text{and}
		\\
		S
		 & = B - \hat B.
		\intertext{Then}
		\left\|S\right\|
		 & \leq
		\frac{2 (\lambda + \delta)}{
			\sigma (\lambda + (\sigma - \delta)^+)
		}.
	\end{align*}
\end{lem}

\begin{lem}[High moment of \(S\)]
	\label{lemma:S-high-moment}
	The random matrix \(S = \hat R^\dagger - R^\dagger\) satisfies
	\begin{align*}
		\expect \left\|S\right\|^{2/\alpha}
		 & \leq
		\frac{
			2^{4/\alpha - 1}\expect \delta^{2/\alpha}
		}{
			\lambda^{2/\alpha} \overline \sigma^{2/\alpha}
			(n')^{1/\alpha}
		}
		+
		\frac{
			2^{4/\alpha - 1}
		}{
			\overline \sigma^{2/\alpha}
			(n')^{1/\alpha}
		}.
	\end{align*}
\end{lem}
\begin{skippedproof}
	Applying Lemma~\ref{lemma:perturbed-tikhonov-pseudoinverse} to the facts of the case,
	\begin{align}
		\left\|S\right\|
		 & \leq
		\frac{{2} \del{
				\delta
				+ \lambda
			}}{
			\overline \sigma \sqrt{n'}
			\del{(\overline \sigma \sqrt{n'} - \delta)^+ + \lambda}
		}
		\label{eq:S-general}
		\intertext{As we will see that
			\(\delta \overset{\mathbb{P}}{\longrightarrow} 0\)
			as \(n \to \infty\), the following bound is very conservative for large \(n\), but it is difficult to do better when we need a high moment of \(S\).
		}
		 & \leq
		\frac{{2} \del{
				\delta
				+ \lambda
			}}{
			\lambda \overline \sigma \sqrt{n'}
		}
		\label{eq:S-conservatism}
		\intertext{Using the fact that \((a + b)^p \leq 2^{p - 1} (a ^p + b^p)\) for \(a, b \geq 0\) and \(p \geq 1\),}
		\left\| S \right\|^{2/\alpha}
		 & \leq
		2^{4/\alpha - 1}
		\frac{
			\delta^{2/\alpha} + \lambda ^ {2/\alpha}
		}{
			\lambda^{2/\alpha} \sigma_\text{min}^{2/\alpha}(\Phi)
		}
		\\
		\expect \left\| S \right\|^{2/\alpha}
		 & \leq
		\frac{
			2^{4/\alpha - 1}\expect \delta^{2/\alpha}
		}{
			\lambda^{2/\alpha} \overline \sigma^{2/\alpha}
			(n')^{1/\alpha}
		}
		+
		\frac{
			2^{4/\alpha - 1}
		}{
			\overline \sigma^{2/\alpha}
			(n')^{1/\alpha}
		}
		\label{eq:S-moment}
	\end{align}
\end{skippedproof}

This bound is not sharp enough to control the \(Su\) error term by H\"older's inequality.
Some improvement is possible by dividing and conquering
\(S
= S\bm{1}_{\delta \leq \lambda}
+ S\bm{1}_{\delta > \lambda}\), head and tail.
Restricted to the head, \(S\) satisfies, in the terms of Lemma~\ref{lemma:perturbed-tikhonov-pseudoinverse}, \(\left\|S\right\| \lesssim \frac{\lambda}{\sigma^2}\); restricted to the tail \(S\) satisfies the \emph{a priori} \(S \leq \frac{1}{\sigma}\), which can be combined by H\"older's inequality with the decaying \(\expect \bm_{\delta > \lambda}\).
However since \(\lambda < \sigma\) by hypothesis and we will later prefer \(\lambda \ll \sigma\), it behooves to get tighter control of \(S\) on the event \(\cbr{\lambda < \delta \leq \sigma}\).

We first state a self-contained technical lemma, proved in Appendix~\ref{section:proof-of-head-body-tail}, concerning Tikhonov pseudoinverses of perturbed matrices.
\begin{lem}[Head-body-tail]
	\label{lemma:head-body-tail}
	Let \(A \in \mathbb{R}^{m\times n}\) have full column rank and minimal singular value \(\sigma\), and let \(D\) be a real random matrix of the same dimensions as \(A\),
	having operator norm \(\delta\).
	For \(\lambda \in (0, \sigma)\), let
	\begin{align*}
		B
		 & = \begin{pmatrix}
			     0_{m\times m} \\ A
		     \end{pmatrix}^\dagger,
		\quad
		\hat B
		=
		\begin{pmatrix}
			\lambda I_m \\ A + D
		\end{pmatrix}^\dagger
		, \quad \text{and}
		\\
		S
		 & = B - \hat B.
		\intertext{Then}
		\expect \left\|S\right\|
		 & \leq s_\text{head}
		+ s_\text{body}
		+ s_\text{tail},
		\intertext{where}
		s_\text{head}
		 & =
		\frac{4\lambda}{\sigma^2},
		\\
		s_\text{body}
		 & =
		3
		\int_\lambda^\sigma
		\frac{(\lambda + \sigma)
			\probability(\delta > v)
		}{
			\sigma (\lambda + \sigma - v)^2
		}
		\dif v
		,
		\shortintertext{and}
		s_\text{tail}
		 & =
		\expect \frac{4\delta}{
			\sigma \lambda
		}
		\bm{1}_{\delta > \sigma}.
	\end{align*}
\end{lem}

Now we return the problem at hand by going through the steps of Lemma~\ref{lemma:head-body-tail}, in which we have an integrability condition on \(\delta\).
\begin{lem}
	\label{lemma:S-first-moment}
	The random matrix \(\left\|S\right\|\) satisfies
	\begin{align*}
		\expect \left\|S\right\|
		 & \leq s_\text{head} + s_\text{body} + s_\text{tail},
		\shortintertext{where}
		s_\text{head}
		 & =
		\frac{4 \lambda}{
			\overline \sigma^2 n'
		},
		\\
		s_\text{body}
		 & =
		\frac{
			6 C_\phi^{2} \epsilon_N^\alpha
		}{
			n^{\frac{2\alpha}{2m + 3}}
		}
		\sbr{
			\frac{
				2 \log (\overline \sigma \sqrt{n'}/\lambda)
			}{
				\overline \sigma^3 \sqrt{n'}
			}
			+ \frac{
				1
			}{
				\lambda \overline \sigma^2
			}
		},
		\shortintertext{and}
		s_\text{tail}
		 & = \frac{
			4
			C_\phi^{2/\alpha} \epsilon_N
		}{
			\lambda \overline \sigma^{2/\alpha}
			n^{\frac{2}{2m + 3}}
		}.
	\end{align*}
\end{lem}
\begin{skippedproof}
	We can immediately transcribe \(s_\text{head}\) from the statement of Lemma~\ref{lemma:head-body-tail}.

	We proceed with \(s_\text{body}\), using Chebyshev's inequality to bound \(\probability(\delta > v)\).
	Using \(\sigma = \overline\sigma \sqrt{n'}\) for temporary brevity,
	\begin{align*}
		 &
		3
		\int_\lambda^\sigma
		\frac{
			(\lambda + \sigma)
			\probability(\delta > v)
		}{
			\sigma (\lambda + \sigma - v)^2
		}
		\dif v
		\\
		 & \leq
		3 \expect \delta^2
		\int_\lambda^\sigma
		\frac{
			\lambda + \sigma
		}{
			(\lambda + \sigma - v)^2 v^2
			\dif v
		}
		\\
		 & =
		\frac{6 \expect \delta^2}{\sigma}
		\sbr{
			\frac{
				2 \log (\sigma/\lambda)
			}{
				(\sigma + \lambda)^2
			}
			+ \frac{
				\sigma - \lambda
			}{
				(\sigma + \lambda) \sigma \lambda
			}
		}
		\\
		\intertext{``Rounding up'' for simplicity,}
		 & \leq
		\frac{6 \expect \delta^2}{\sigma}
		\sbr{
			\frac{
				2 \log (\sigma/\lambda)
			}{
				\sigma^2
			}
			+ \frac{
				1
			}{
				\sigma \lambda
			}
		}
		\intertext{Applying Lemma~\ref{lemma:delta-moment} we get this lemma's \(s_\text{body}\).}
		 & \leq
		\frac{
			6 C_\phi^{2} \epsilon_N^\alpha
		}{
			n^{\frac{2\alpha}{2m + 3}}
		}
		\sbr{
			\frac{
				2 \log (\overline \sigma \sqrt{n'}/\lambda)
			}{
				\overline \sigma^3 \sqrt{n'}
			}
			+ \frac{
				1
			}{
				\lambda \overline \sigma^2
			}
		}.
	\end{align*}

	As for \(s_\text{tail}\), by H\"older's inequality,
	\begin{align*}
		 & \expect \frac{4\delta}{
			\sigma \lambda
		}
		\bm{1}_{\delta > \sigma}
		\\
		 & \leq
		\frac{
			4
		}{
			\lambda \overline \sigma \sqrt{n'}
		}
		\del{
			\expect \delta^{2/\alpha}
		}^{\alpha/2}
		\sbr{
			\probability\del{\delta > \overline \sigma \sqrt{n'}}
		}^{(2 - \alpha)/2}
		\\
		 & \leq
		\frac{
			4
		}{
			\lambda \overline \sigma \sqrt{n'}
		}
		\del{
			\expect \delta^{2/\alpha}
		}^{\alpha/2}
		\sbr{
			\frac{
				\expect \delta^{2/\alpha}
			}{
				(\overline \sigma^2 n')^{1/\alpha}
			}
		}^{(2 - \alpha)/2}
		\\
		 & =
		\frac{
			4
		}{
			\lambda \overline \sigma \sqrt{n'}
		}
		\frac{
			\expect \delta^{2/\alpha}
		}{
			(\overline \sigma^2 n')^{
					\frac{2 - \alpha}{2\alpha}
				}
		}
		\\
		 & =
		\frac{
			4
			\expect \delta^{2/\alpha}
		}{
			\lambda \overline \sigma^{2/\alpha}
			(n')^{1/\alpha}
		}
		\\
		 & \leq
		\frac{
			4
			C_\phi^{2/\alpha} \epsilon_N
		}{
			\lambda \overline \sigma^{2/\alpha}
			n^{\frac{2}{2m + 3}}
		}.
	\end{align*}
\end{skippedproof}

\subsection{Bound on \(\Delta \theta\)}
We have now controlled each term in \eqref{eq:Delta-theta},
\begin{align*}
	\left\|\Delta \theta\right\|
	 & \leq \frac{\left\|v\right\|}{\sigma_\text{min} (R)}
	+ \left\|S\right\| \left\|u\right\| +
	\left\|S\right\| \left\|v\right\|.
\end{align*}
\begin{thm}[First moment of \(\Delta \theta\)]
	\label{thm:delta-theta-moment}
	The random vector \(\Delta \theta\) satisfies
	\begin{align*}
	\begin{split}
		\expect \left\|\Delta \theta\right\|
		 &\leq
		\frac{
			\epsilon_{N, m}^{1/2}
		}{
			\overline \sigma n^{\frac{1}{2m + 3}}
		}
		+
		\frac{4 \bar U  \lambda}{
			\overline \sigma^2 \sqrt{n'}
		}
		\\
		&\quad+
		\frac{
			6 C_\phi^{2\alpha} \bar U \epsilon_N^\alpha
		}{
			n^{\frac{2\alpha}{2m + 3}}
		}
		\sbr{
			\frac{
				2 \log (\overline \sigma \sqrt{n'}/\lambda)
			}{
				\overline \sigma^3
			}
			+ \frac{
				\sqrt{n'}
			}{
				\lambda \overline \sigma^2
			}
		}
		\\
		&\quad+
		\frac{
			4
			C_\phi^2 \bar U \epsilon_N \sqrt{n'}
		}{
			\lambda \overline \sigma^{2/\alpha}
			n^{\frac{2}{2m + 3}}
		}
		\\
		&\quad+
		\frac{
			2^{2 - \frac{\alpha}{2}}
			\epsilon_{N, m}^{1/2}
		}
		{
			n^{\frac{1}{2m + 3}}
		}
		\del{
			\frac{
				C_\phi^2 \epsilon_N (n')^{1/\alpha}
			}{
				\overline \sigma^{2/\alpha} n^{\frac{2}{2m + 3} }
				\lambda^{2/\alpha}
			}
			+
			\frac{
				1
			}{
				\overline \sigma^{2/\alpha}
			}
		}^{\alpha/2}.
		\end{split}
	\end{align*}
\end{thm}
\begin{skippedproof}
	According to Lemma~\ref{lemma:v-moment},
	\begin{align*}
		\expect \frac{\left\|v\right\|}{\sigma_\text{min}(R)}
		 & \leq
		\frac{
			\epsilon_{N, m}^{1/2}
		}{
			\overline \sigma n^{\frac{1}{2m + 3}}
		}
	\end{align*}

	According to Lemma~\ref{lemma:S-first-moment},
	we can bound \(\left\|u\right\| \expect \left\|S\right\|\) as the sum of
	\begin{align*}
		\left\|u\right\| s_\text{head}
		 & =
		\frac{4 \bar U  \lambda}{
			\overline \sigma^2 \sqrt{n'}
		},
		\\
		\left\|u\right\| s_\text{body}
		 & =
		\frac{
			6 C_\phi^{2\alpha} \bar U \epsilon_N^\alpha
		}{
			n^{\frac{2\alpha}{2m + 3}}
		}
		\sbr{
			\frac{
				2 \log (\overline \sigma \sqrt{n'}/\lambda)
			}{
				\overline \sigma^3
			}
			+ \frac{
				\sqrt{n'}
			}{
				\lambda \overline \sigma^2
			}
		},
		\shortintertext{and}
		\left\|u\right\| s_\text{tail}
		 & = \frac{
			4
			C_\phi^2 \bar U \epsilon_N \sqrt{n'}
		}{
			\lambda \overline \sigma^{2/\alpha}
			n^{\frac{2}{2m + 3}}
		}.
	\end{align*}

	Using Lemmas~\ref{lemma:S-high-moment} and \ref{lemma:v-moment} in H\"older's inequality,
	\begin{align*}
		\expect \left\|S\right\| \left\|v\right\|
		 & \leq
		\del{
		\expect \left\|S\right\|^{2/\alpha}
		}^{\alpha/2}
		\del{
		\expect \left\|v\right\|^{2/(2 - \alpha)}
		}^{(2 - \alpha)/2}
		\\
		 & \leq
		\del{
			\frac{
				2^{4/\alpha - 1}\expect \delta^{2/\alpha}
			}{
				\lambda^{2/\alpha} \overline \sigma^{2/\alpha}
				(n')^{1/\alpha}
			}
			+
			\frac{
				2^{4/\alpha - 1}
			}{
				\overline \sigma^{2/\alpha}
				(n')^{1/\alpha}
			}
		}^{\alpha/2}
		\notag
		\\
		 & \quad
		\cdot
		\del{
			\frac{
				\epsilon_{N, m}^{1/2}
				(n')^{1/2}
			}
			{
				n^{\frac{1}{2m + 3}}
			}
		}
		\notag
		\\
		 & \leq
		\frac{
			2^{2 - \frac{\alpha}{2}}
			\epsilon_{N, m}^{1/2}
		}
		{
			n^{\frac{1}{2m + 3}}
		}
		\del{
			\frac{
				\expect \delta^{2/\alpha}
			}{
				\lambda^{2/\alpha} \overline \sigma^{2/\alpha}
			}
			+
			\frac{
				1
			}{
				\overline \sigma^{2/\alpha}
			}
		}^{\alpha/2}
		\\
		 & \leq
		\frac{
			2^{2 - \frac{\alpha}{2}}
			\epsilon_{N, m}^{1/2}
		}
		{
			n^{\frac{1}{2m + 3}}
		}
		\del{
			\frac{
				C_\phi^{2/\alpha} \epsilon_N (n')^{1/\alpha}
			}{
				\lambda^{2/\alpha}
				\overline \sigma^{2/\alpha} n^{\frac{2}{2m + 3} }
			}
			+
			\frac{
				1
			}{
				\overline \sigma^{2/\alpha}
			}
		}^{\alpha/2}.
	\end{align*}
\end{skippedproof}
The terms in this bound go both ways with respect to \(n'\).
On one hand, there is \((n')^{-1/2}\) scaling reminiscent of the concentration of measure in conventional linear regression.
On the other hand, pseudo-inverting a random matrix with \(n'\) rows offers \(n'\) ways to go wrong, and we get a \((n')^{1/2}\) scaling in those terms.

A parsimonious choice of \(\lambda\) is
\begin{align}
	\frac{\lambda^*}{\sqrt{n'}}
	 &
	=
	\frac{
		C_\phi \epsilon_N^{\alpha/2}
	}{
		n^{\alpha/(2m + 3)}
	},
	\label{eq:lambda-star}
\end{align}
which consolidates the two terms of Lemma~\ref{lemma:S-high-moment}, making the \(\left\|S\right\| \left\|v\right\|\) estimate proportional to the \(\left\|v\right\|/\sigma_\text{min}(R)\) estimate.
Applying the choice \(\lambda = \lambda^*\) of \eqref{eq:lambda-star} to Thm.~\ref{thm:delta-theta-moment} proves Theorem~\ref{thm:mae-after-lambda}.
	
	\section{Proof of Lemma~\ref{lemma:perturbed-tikhonov-pseudoinverse}}
\label{appendix-proof:perturbed-tikhonov-pseudoinverse}
Using Thm.~4.1 of \cite{wedin_perturbation_1973} for the full-rank perturbation of a full-rank rectangular matrix,
\begin{align}
	\left\|S\right\|
	 & \leq
	\sqrt{2} \left\|
	\begin{pmatrix}
		0_{n\times n} \\ A
	\end{pmatrix}^\dagger
	\right\|
	\left\|
	\begin{pmatrix}
		\lambda I_n \\ A + D
	\end{pmatrix}^\dagger
	\right\|
	\left\|
	\begin{pmatrix}
		\lambda I_n \\
		D
	\end{pmatrix}
	\right\|.
	\label{eq:a-priori-stewart}
\end{align}
Recall that the operator norm of a matrix's pseudoinverse is the inverse of its least singular value. Thus,
\begin{align}
	\left\|
	\begin{pmatrix}
		0_{n\times n} \\ A
	\end{pmatrix}^\dagger
	\right\|
	 & = \frac{1}{\sigma}
	\intertext{and}
	\left\|
	\begin{pmatrix}
		\lambda I_n \\ A + D
	\end{pmatrix}^\dagger
	\right\|^{-2}
	 & = \lambda^2 + \sigma_\text{min}^2(A + D).
	\notag
	\\
	\intertext{By Jensen's and Weyl's inequalities,}
	\left\|
	\begin{pmatrix}
		\lambda I_n \\ A + D
	\end{pmatrix}^\dagger
	\right\|^{-1}
	 & \geq
	\frac{
		\lambda + \sigma_\text{min}(A + D)
	}{
		\sqrt 2
	}.
	\notag
	\\
	 & \geq
	\frac{
		\lambda + (\sigma - \delta)^+
	}{
		\sqrt 2
	}
\end{align}
where \(\del{\cdot}^+ = \max(0, \cdot)\). We bound the last factor of \eqref{eq:a-priori-stewart} using the triangle inequality.%
}{}
\section{Proof of Lemma~\ref{lemma:head-body-tail}}
\label{section:proof-of-head-body-tail}
Recall the \emph{a priori} bound from Lemma~\ref{lemma:perturbed-tikhonov-pseudoinverse},
\begin{align}
	\left\|S\right\|
	 & \leq
	\frac{2(\delta + \lambda)}{
		\sigma (\lambda + (\sigma - \delta)^+)
	}
	\label{eq:proof-head-body-tail-a-priori}.
\end{align}
We will integrate \(\expect \left\|S\right\|\) over three zones of \(\delta\): ``head,'' \([0, \lambda]\); ``body,'' \((\lambda, \sigma]\); and ``tail,'' \((\sigma, \infty)\).

First, we have the deterministic
\begin{align}
	\left\|S\right\| \bm{1}_{\delta \leq \lambda}
	 & \leq
	\frac{4\lambda}{
		\sigma^2
	}\bm{1}_{\delta \leq \lambda}
	\label{eq:head-body-tail-first-raw}
\end{align}

Second, define the function \(F(x) = \frac{
	2(\lambda + x)
}{
	\sigma(\lambda + \sigma - x)
}\).
We will execute a ``layer cake'' rearrangement of \(F(\delta)\).
\begin{align}
	\left\|S\right\| \bm{1}_{\lambda < \delta \leq \sigma}
	 & \leq
	F(\delta)
	\bm{1}_{\lambda < \delta \leq \sigma}
	\notag
	\\
	\intertext{By the Fundamental Theorem of Calculus,}
	 & =
	\del{
		F(\lambda)
		+ \int_\lambda^\delta
		F'(v)
		\dif v
	}
	\bm{1}_{\lambda < \delta \leq \sigma}
	\\
	 & =
	F(\lambda) \bm{1}_{\lambda < \delta}
	+
	\del{
		\int_\lambda^\delta
		F'(v)
		\dif v
	}
	\bm{1}_{\lambda < \delta \leq \sigma}
	\label{eq:head-body-tail-second-raw}
\end{align}

Let us combine the former term of \eqref{eq:head-body-tail-second-raw} with \eqref{eq:head-body-tail-first-raw}, and call it \(s_\text{head}\).
The latter term we will call \(s_{\text{body}, 0}\).
We get the bound
\begin{align*}
	\left\|S\right\|\bm{1}_{\delta \leq \sigma}
	 & \leq s_\text{head} + s_{\text{body}, 0},
	\\
	s_{\text{body}, 0}
	 & =
	\del{
		\int_\lambda^\delta
		F'(v)
		\dif v
	}
	\bm{1}_{\lambda < \delta \leq \sigma}
	\\
	 & =
	\del{
		\int_0^\infty
		F'(v)
		\bm{1}_{\lambda < v < \delta}
		\dif v
	}
	\bm{1}_{\lambda < \delta \leq \sigma}
	\\
	\shortintertext{We conservatively (giving up the predicate \(\delta \leq \sigma\)) rework the logic so that the bounds of integration do not depend on \(\delta\).}
	 & \leq
	\int_0^\infty
	F'(v)
	\bm{1}_{\lambda < v < \sigma}
	\bm{1}_{v < \delta}
	\dif v
	\\
	 & =
	\int_\lambda^\sigma
	F'(v)
	\bm{1}_{v < \delta}
	\dif v
	\shortintertext{Taking the expectation of both sides and applying Tonelli's theorem,}
	\expect s_{\text{body},0}
	 & \leq
	\int_\lambda^\sigma
	F'(v)
	\probability(\delta > v)
	\dif v
	\\
	\shortintertext{By Chebyshev's inequality,}
	 & =
	\int_\lambda^\sigma
	\frac{
		2(2\lambda + \sigma)
		\probability(\delta > v)
	}{
		\sigma (\lambda + \sigma - v)^2
	}
	\dif v
	\shortintertext{To balance \(\lambda\) and \(\sigma\), we trade in \(\lambda/2\) for \(\sigma/2\).}
	 & \leq
	3
	\int_\lambda^\sigma
	\frac{(\lambda + \sigma)
		\probability(\delta > v)
	}{
		\sigma (\lambda + \sigma - v)^2
	}
	\dif v
\end{align*}
We call this last quantity \(s_\text{body}\).
When \(\delta\) is large, \eqref{eq:proof-head-body-tail-a-priori} becomes
\begin{align*}
	\left\|S\right\| \bm{1}_{\delta > \sigma}
	 & \leq
	\frac{3(\delta + \lambda)}{
		\sigma \lambda
	}
	\bm{1}_{\delta > \sigma}.
	\shortintertext{We do not lose much by simplifying \(\lambda < \sigma < \delta\).}
	 & \leq
	\frac{4\delta}{
		\sigma \lambda
	}
	\bm{1}_{\delta > \sigma}.
\end{align*}
The expectation of this quantity constitutes \(s_\text{tail}\).

\end{document}